\documentclass{epl}
\newcommand{\be}{\begin{equation}}
\newcommand{\bea}{\begin{eqnarray}}
\renewcommand{\d}{{\rm d}}
\newcommand{\ee}{\end{equation}}
\newcommand{\eea}{\end{eqnarray}}
\newcommand{\mean}[1]{\left\langle#1\right\rangle}

\newcommand{\eff}{_{{\rm eff}}}
\newcommand{\vd}{{\bf d}}
\renewcommand{\c}{{\bf c}}
\newcommand{\e}{{\bf e}}
\renewcommand{\k}{{\bf k}}
\newcommand{\x}{{\bf x}}
\newcommand{\y}{{\bf y}}
\newcommand{\0}{{\bf 0}}
\newcommand{\V}{{\bf V}}
\newcommand{\p}{\partial}
\newcommand{\dt}{\delta t}

\newcommand{\frah}[2]{#1/#2}

\title{Anomalous aging phenomena caused by drift velocities}
\shorttitle{Anomalous aging phenomena}
\author{J.M.~Luck\inst{1}\footnote{E-mail: luck@spht.saclay.cea.fr}
\and Anita Mehta\inst{2}\footnote{E-mail: anita@boson.bose.res.in}}
\institute{
\inst{1}Service de Physique Th\'eorique,
CEA Saclay, 91191 Gif-sur-Yvette cedex, France\\
\inst{2}ICTP, Strada Costiera 11, 34100 Trieste, Italy,
and S N Bose National Centre for Basic Sciences, Block JD Sector 3,
Salt Lake, Calcutta 700098, India}
\begin{document}
\maketitle
\begin{abstract}
We demonstrate via several examples that a uniform drift velocity
gives rise to anomalous aging,
characterized by a specific form for the two-time correlation functions,
in a variety of statistical-mechanical systems far from equilibrium.
Our first example concerns the oscillatory phase
observed recently in a model of competitive learning.
Further examples, where the proposed theory is exact,
include the voter model and the Ohta-Jasnow-Kawasaki theory for domain growth
in any dimension, and a theory for the smoothing of sandpile surfaces.
\end{abstract}

\def\fun{
\begin{figure}
\vskip 7cm{\hskip 1.4cm}
\includegraphics{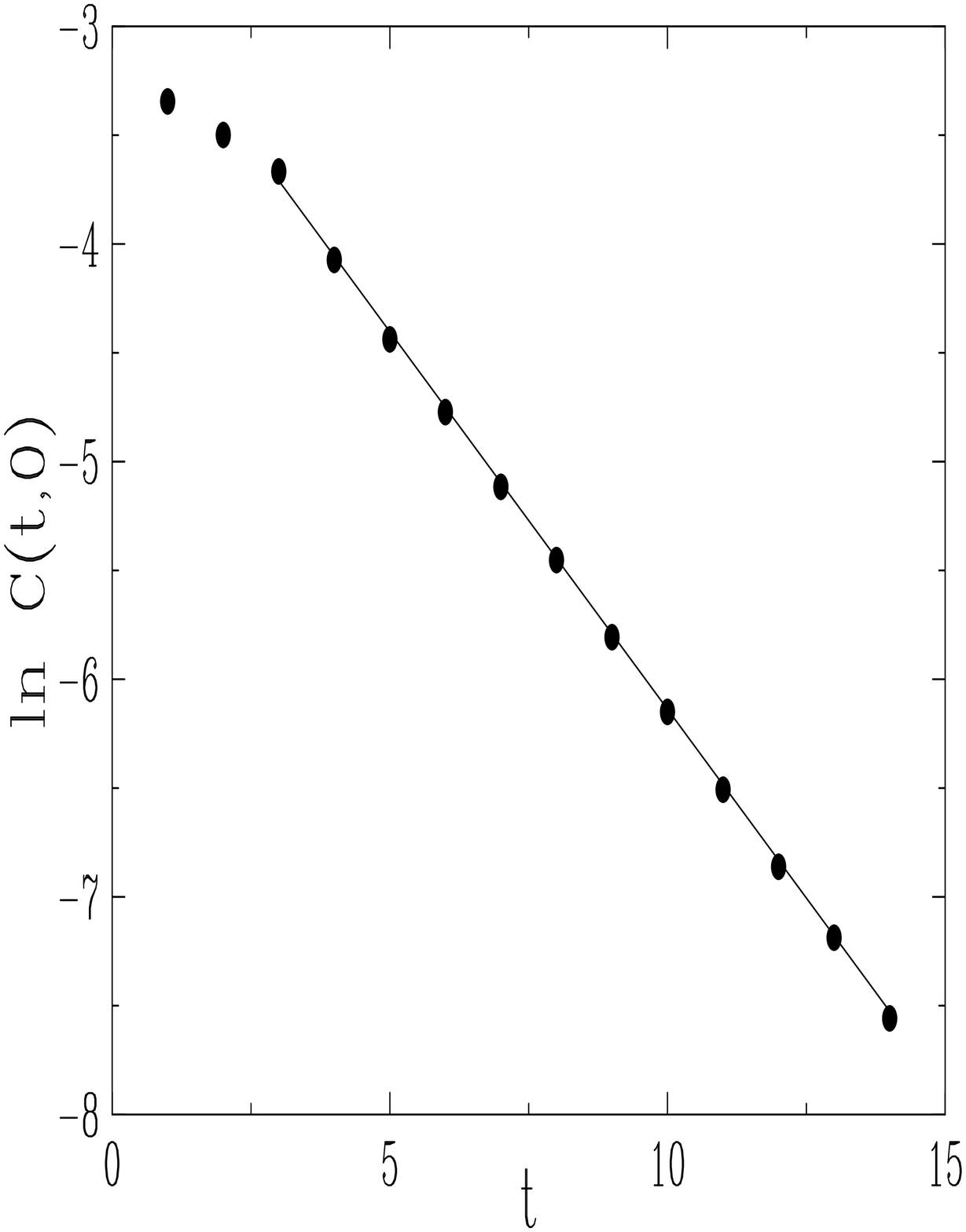}
\caption{Logarithmic plot of efficiency correlation function $C(t,0)$
against $t$.
Full line: least-square fit of data for $t\ge3$ (slope $=-0.34$).}
\label{fun}
\end{figure}}

\def\fde{
\begin{figure}
\vskip 7cm{\hskip 1.4cm}
\includegraphics{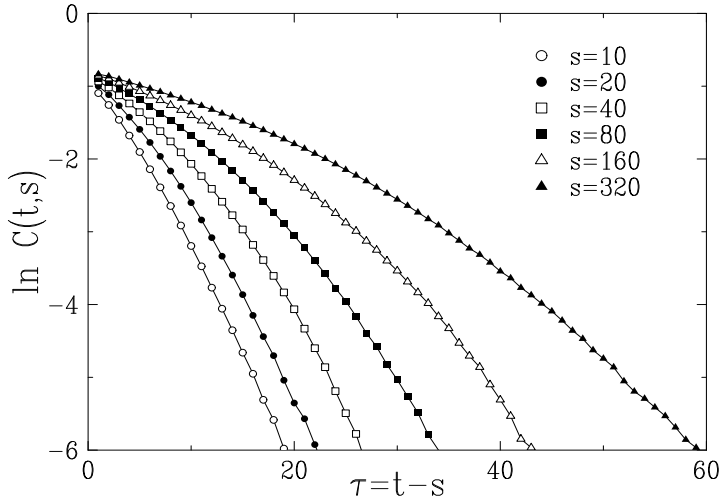}
\caption{Logarithmic plot of efficiency correlation function $C(t,s)$
against $\tau=t-s$, for several values of the waiting time~$s$.}
\label{fde}
\end{figure}}

\def\ftr{
\begin{figure}
\vskip 7cm{\hskip 1.4cm}
\includegraphics{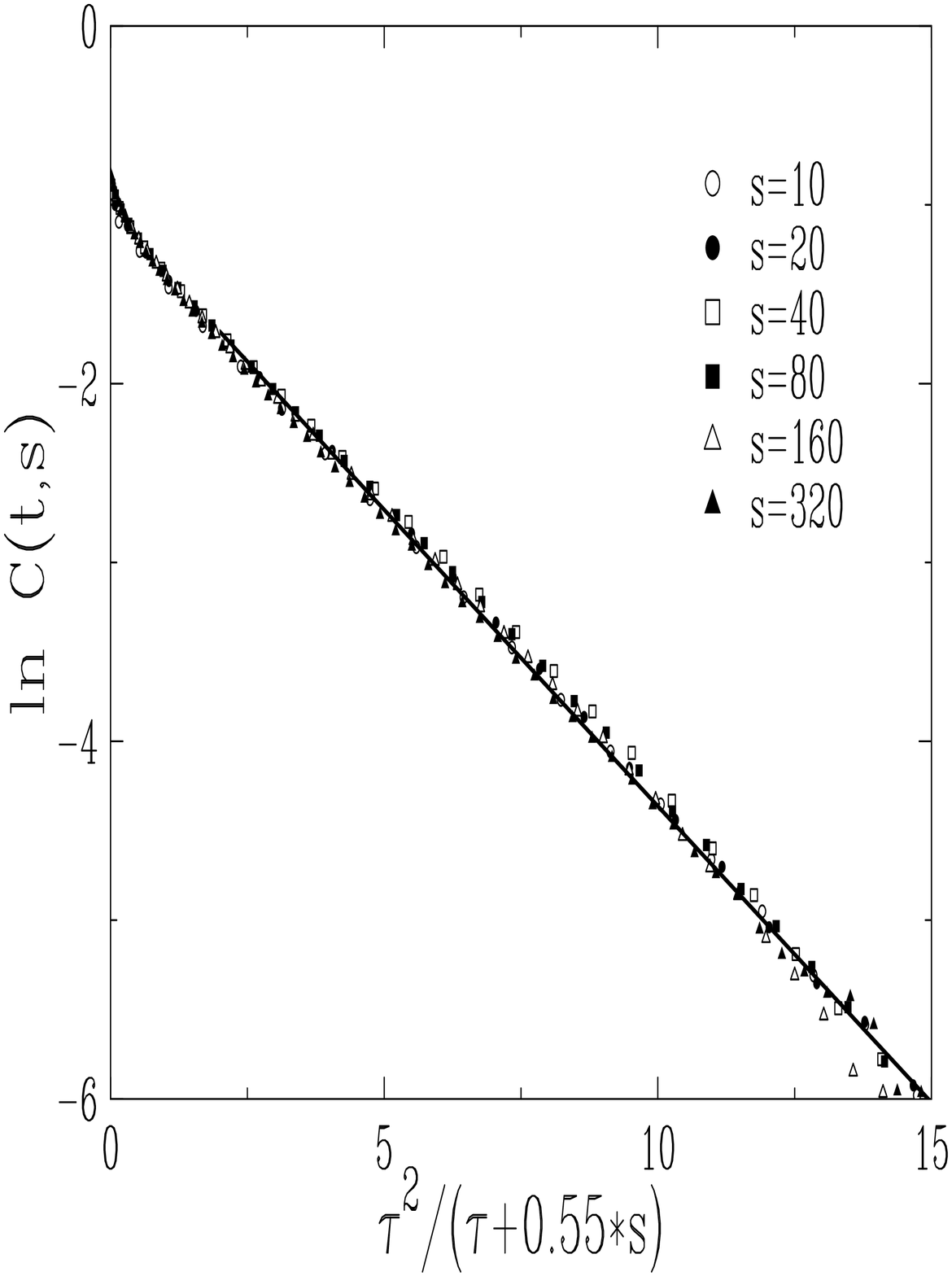}
\caption{Logarithmic plot of the data of Figure~\ref{fde},
against $\tau^2/(\tau+0.55\,s)$.
Full line: least-square fit of data with abscissa $>2$ (slope $=-0.33$).}
\label{ftr}
\end{figure}}

Aging is a characteristic feature of systems
such as glasses and spin glasses in their low-temperature phase.
More generally, aging is a consequence of slow dynamics,
which can be observed in a broad range of systems
far from equilibrium~\cite{angell}.
Aging phenomena are commonly described in terms of the two-time
correlation and response functions of the relevant variables,
which often exhibit scaling behavior in the two-time plane~\cite{aging}.

In this Letter we demonstrate that this aging scenario is strongly affected
by the presence of a uniform drift velocity,
either intrinsic to the microscopic dynamics,
or imposed by external conditions.
On general grounds, a uniform velocity field is expected
to be relevant for long-time dynamics.
Consider for simplicity a diffusive relaxational dynamics,
characterized by a diffusion constant $D$.
To a non-zero velocity $V$ naturally corresponds a finite relaxation
time,~$D/V^2$, beyond which the dynamical evolution
can be qualitatively different from that with~$V=0$.

Non-equilibrium phenomena in the presence of velocity fields,
either uniform (drift) or non-uniform (shear),
and more generally in a range of situations with a preferred direction,
have been described in recent works.
For instance, it has been shown that a shear flow
induces a strong anisotropy in coarsening dynamics~\cite{bc}.
The influence of drift on phase-ordering dyna\-mics
has also been investigated in Ref.~\cite{brag}.
The peculiar scaling properties of the flocking model of Ref.~\cite{ole}
are yet another consequence of directionality in non-equilibrium systems.

Here we show that a variety of models
exhibit a novel type of non-equilibrium dynamics
in the presence of a uniform drift velocity $V$,
characterized by the specific scaling form~(\ref{th})
for the two-time correlation function.
We refer to this phenomenon as anomalous aging.
Uniform drift velocities cannot here be transformed away naively,
by invoking Galilean invariance.

We first investigate in detail
the oscillatory phase described recently in the context of a model
of competitive learning~\cite{pre}.
The anomalous aging behavior of this specific model
will be explained at a quantitative level
by the combined effects of drift and diffusion.
Anomalous aging will then be shown to be a common feature
of a range of non-equilibrium models in the presence of a drift,
including the voter model, the OJK theory for domain growth,
and the smoothing of sandpile surfaces~\cite{mln,coupled}.

We begin with a brief description of our model
of competitive learning~\cite{pre}.
Two types of individuals, considered as quick and slow learners,
are respectively classified by their efficiencies
(coded as Ising variables $\eta_\x=\pm1$),
and occupy the sites of a regular lattice.
The updating dynamics of the efficiencies involves two steps.
First, a majority rule favors the conversion of every individual
to the dominant type among its neighbors.
In the next step, there is a majority rule
involving the outcomes of individuals in some stochastic game,
favoring the conversion of every individual to the
type which has had the most favorable outcomes among its neighbors
at the last iteration.
Our model has two variants, the `interfacial' and the `cooperative',
which differ in the precise formulation of the second step.

We consider the situation where both types
of individuals, while distinguishable, are equally slow learners,
so that no type wins {\it a priori}.
We refer to this case as coexistence, in analogy with ferromagnetic models.
Several kinds of ordered and disordered phases,
i.e., types of non-equilibrium steady states,
have been found in the phase diagram of both variants of the model~\cite{pre};
most phase transitions were found to belong either to the Ising
or to the voter universality classes.
However, an unusual type of coarsening was found in the
cooperative version of the model,
where the order parameter underwent more or less regular oscillations in time,
hence the name {\it oscillatory phase}.

As the model of competitive learning is entirely defined by dynamical rules,
without reference to a Hamiltonian
(and therefore without the concept of detailed balance),
details of the dynamics can influence its steady states.
It turns out that
the updating scheme, including the order in which sites are updated
(sequential or parallel, ordered or random), {\em does}
affect the phase diagram.
From now on, we focus on the cooperative model at coexistence
on the square lattice,
characterized by a single parameter, $0\le p\le1$,
which represents an individual's probability of obtaining a successful outcome.
In Ref.~\cite{pre} ordered sequential dynamics were used,
and an oscillatory phase was found for $p<p_o$ $(p_o\approx0.136)$.
On the other hand,
parallel dynamics leads to a ferromagnetically ordered phase
for $p<p_c$ $(p_c\approx0.259)$,
while random sequential dynamics leads to a disordered phase
with very weak correlations between efficiencies.

The special feature of ordered sequential dynamics
that is responsible for the oscillatory phase
turns out to be its {\it directionality}.
Consider a square sample with sites $\x=(m,n)$ with $m,n=1,\dots,L$.
For ordered sequential dynamics, neglecting boundary effects,
the next site updated after $\x_t=(m,n)$ is
$\x_{t+\dt}=(m,n+1)$, with $\dt=1/L^2$.
We define the directionality of this updating scheme as
$\vd=\mean{\x_{t+\dt}-\x_t}$.
We have $\vd=(0,1)$ for ordered sequential dynamics,
while $\vd=\0$ for parallel dynamics or random sequential dynamics.
Other updating schemes have been considered,
including alternating sequential dynamics:
the sample is first swept in a regular way,
such as e.g.~$\x_{t+\dt}=(m,n+1)$ if $\x_t=(m,n)$,
and then in a different regular way,
such as $\x_{t+\dt}=(m,n-1)$, or $\x_{t+\dt}=(m+1,n)$.
We have $\vd=\0$ in the first example,
and $\vd=\frac12(1,1)\ne\0$ in the second.
The model exhibits an oscillatory phase
at low~$p$ if, and only if, the updating scheme has
a net directionality $\vd\ne\0$.

In the following, we restrict the analysis to the prototypical example
of ordered sequential dynamics.
Spin (efficiency) patterns are advected with a constant drift velocity $\V$,
due to the directionality $\vd$ of the dynamics.
The direction of $\V$ is given by the arrow of $\vd$,
while its strength $V$ depends e.g. on the parameter $p$.
If periodic boundary conditions are used, spin clusters wrap around the sample,
so that the magnetization oscillates more or less regularly,
with period $T_1=2L/V$, which defines a time scale associated with
the drift velocity.
This phenomenon was depicted in Figure~11 of Ref.~\cite{pre}.
A correlation analysis of a long magnetization record
allows for the accurate determination of $V$.
We find that $V$ decreases roughly linearly with $p$,
from $V\approx1.14$ at $p=0$ to $V\approx0.97$ at $p=p_o\approx0.136$.
The drift velocity
thus remains of order unity throughout the oscillatory phase,
including the critical point $p_o$,
where it does not seem to exhibit any kind of critical behavior.

Next, a coarsening phenomenon takes place
if the system starts from a random initial state
with uncorrelated efficiencies.
The commonly accepted picture of phase-ordering dynamics
with no conservation law~\cite{bray} involves diffusive behavior,
hence the emergence of a second time scale, $T_2=L^2/D$,
with the diffusion coefficient $D$ being a microscopic constant.
The competition between these two time scales will turn out to be central
to anomalous aging [see paragraph below Eq.~(\ref{taueff})].

Along the lines of investigations of aging properties in glassy systems,
we next measured the two-time correlation function
of the efficiency of a given individual,
$C(t,s)=\mean{\eta_\x(s)\eta_\x(t)}$,
with $0\le s$ (waiting time) $\le t=s+\tau$ (observation time).
We have chosen to work at $p=0$ for definiteness.
For each value of the waiting time $s$,
data were averaged over more than 1,000 samples
of size $300\times 300$.
The particular case $s=0$ corresponds to the correlation
of the efficiencies at time $t$ with their random initial values.
Figures~\ref{fun} and~\ref{fde} respectively show plots
of $C(t,0)$ and $C(t,s)$ for $s>0$.
Figure~\ref{fun} demonstrates that $C(t,0)$ decays exponentially,
just as in any equilibrium process,
with a microscopic relaxation time $\tau_0\approx1/0.34\approx2.9$,
while Figure~\ref{fde} clearly shows non-stationarity or aging effects:
correlations decay more and more slowly when the waiting time is increased.

\fun

\fde

In order to elucidate these observations,
and especially to understand better the role of the drift velocity $V$
on the non-equilibrium dynamics of the model,
we propose the following phenomenological linear theory
meant to incorporate the combined effects of drift and diffusion,
in a natural fashion and at a minimal level of sophistication.
The efficiencies $\eta_\x$ are replaced by a continuous field $\phi(\x,t)$.
The equal-time correlation function $C(\x,t)=\mean{\phi(\x,t)\phi(\0,t)}$
is assumed to obey the diffusion equation
\be
\frah{\p C(\x,t)}{\p t}=D_1\nabla^2C(\x,t),
\label{eq1}
\ee
as the drift velocity is expected to have no effect on one-time observables.
A disordered initial state corresponds to
$C(\x,0)=\delta^{(d)}(\x)$ at $t=0$, in dimension $d$.
The two-time correlation function
$C(\x,t,s)=\mean{\phi(\x,t)\phi(\0,s)}$ is assumed
to be affected by both the drift and the diffusion mechanisms:
\be
\frah{\p C(\x,t,s)}{\p t}=D_2\nabla^2C(\x,t,s)-\V\cdot\nabla C(\x,t,s)
\label{eq2}
\ee
$(t\ge s)$, with initial condition $C(\x,s,s)=C(\x,s)$ at $t=s$.
The diffusion coefficients $D_1$ and $D_2$ are possibly different.
Convoluting the Green's functions of both equations,
and neglecting pre-exponential factors, we obtain
\be
C(t=s+\tau,s)\approx\exp\left(-\frac{V^2\tau^2}{4(D_1s+D_2\tau)}\right).
\label{th}
\ee

This analytical result is in excellent agreement with our results above.
First, Eq.~(\ref{th}) predicts that $C(t,0)$ decays exponentially,
with a relaxation time $\tau_0=4D_2/V^2$, as seen in
Figure~\ref{fun}, where
the values of $V$ and $\tau_0$ yield $D_2\approx0.94$.
Second, in order to check the specific form of Eq.~(\ref{th}),
the data for $\ln C(t,s)$ have been plotted
as a function of $\tau^2/(\tau+\alpha s)$.
A convincing data collapse and linear behavior are observed
(Figure~\ref{ftr}) for the optimal value $\alpha\approx0.55$,
yielding $D_1=\alpha D_2\approx0.52$.
Moreover, Eq.~(\ref{th}) predicts that the slopes
of Figures~\ref{fun} and~\ref{ftr} are equal;
the measured slopes of the least-square fits, $-0.33$ and $-0.34$,
coincide within errors, in agreement with the above.

\ftr

Our result~(\ref{th}) is the centerpiece of {\it anomalous aging}.
The correlation function therein is clearly aging,
in the general sense that it is not stationary,
and decays more and more slowly as the waiting time $s$ is increased.
The expression~(\ref{th}) does not, however,
exhibit any simple scaling form as a function of the time ratio $\tau/s$,
since it involves the intrinsic time scale $\tau_0=4D_2/V^2$,
emblematic of the competition between drift and diffusion,
as expected from the discussion in the beginning of this Letter.

For $\tau\ll s$,
the correlation function in Eq.~(\ref{th}) has a Gaussian scaling form
\be
C(t,s)\approx\exp\big(-\tau^2/(\tau_1s)\big)\qquad(\tau=t-s\ll s),
\label{bulk}
\ee
with $\tau_1=4D_1/V^2=\tau_0D_1/D_2$.
The correlation is characterized by an effective relaxation time
\be
\tau\eff(s)\approx(\tau_1s)^{1/2},
\label{taueff}
\ee
which diverges, albeit anomalously slowly, with the waiting time $s$.

Eq.~(\ref{taueff}) can be recast as $(V\tau\eff)^2\approx4D_1s$,
to be put in perspective with the similar quadratic relationship
$(VT_1)^2=4DT_2=4L^2$ which holds between both time scales introduced above.
The stationary or time-translationally-invariant (TTI) regime
usually observed for $\tau\ll s$,
where the correlation function only depends on $\tau$,
is absent in the present case, having already
died away before our continuum description applied.
The TTI regime is also too small to be visible on Figure~\ref{fde}.

For $\tau\gg s$, the result~(\ref{th}) only depends on $\tau$,
and falls off exponentially, as
\be
C(t,s)\approx\exp(-\tau/\tau_0)\qquad(\tau=t-s\gg s),
\label{tail}
\ee
just as $C(t,0)$.
This paradoxical emergence of a translationally-invariant regime
for {\it long} time separations is one of the reasons
we refer to this aging as {\em anomalous}.
The crossover between both kinds of behavior~(\ref{bulk}) and~(\ref{tail})
takes place for $\tau\sim s$,
where the correlation is already exponentially small,
of order $\exp(-s/\tau_0)$.

In agreement with general expectations on aging phenomena~\cite{aging},
the result~(\ref{th}) can be recast as $C(t,s)\approx F[h(t)/h(s)]$,
with $h(t)=\exp\left(2(t/\tau_1)^{1/2}\right)$
and $F(x)=\exp\left(-(\ln x)^2\right)$.
Anomalous aging thus appears as a special case of sub-aging,
using the terminology of Ref.~\cite{aging},
as the effective age function $h(t)$ grows much more rapidly
than real time $t$.
The effective relaxation time~(\ref{taueff}) is recovered as
$\tau\eff(t)\sim h(t)/(\d h(t)/\d t)$.
A similar behavior, with the same age function $h(t)$,
has been found~\cite{let} in the spherical model with random
asymmetric couplings at zero temperature.
The ferromagnetic spherical model with a conserved order parameter~\cite{bepjb}
provides another interesting instance of anomalous aging,
involving a scaling function similar to the above $F(x)$.

Since Eq.~(\ref{th}) is a simple consequence
of the combined effects of drift and diffusion,
it might be expected to apply to a variety of situations.
Several examples are discussed below,
where anomalous aging applies exactly in the regime of long times.

First, our linear theory is exact for the voter model.
The dynamics of this model~\cite{L,vm,vp}
amounts to saying that a given site, say $\x$,
takes the opinion of any of its neighbors, say $\y=\x+\e$,
with the lattice vector $\e$ being chosen at random with some weight $w(\e)$.
If all vectors $\e$ have equal weights
($w(\e)=1/z$ for all $\e$, with $z$ the coordination number),
the usual voter model is recovered,
with its well-known diffusive dynamics~\cite{vm,vp}.
In the general (asymmetric) case where some neighbors
are preferred over others,
patterns of opinions are advected with a velocity $\V=-\sum_\e\e\,w(\e)$.
We have shown that the equal-time and two-time correlation functions
obey Eqs.~(\ref{eq1}) and~(\ref{eq2}) in the continuum limit, with the identity
\be
D_1=2D_2,
\label{iden}
\ee
so that anomalous aging applies exactly to the voter model.

Second, the Ohta-Jasnow-Kawasaki (OJK) theory~\cite{ojk},
a linear approximation commonly used in the study
of phase-ordering dynamics~\cite{bray},
also yields, in the presence of drift, the anomalous aging result~(\ref{th}).
The main difference between the OJK theory and our linear theory
is that linear equations are assumed to be obeyed
by the field $\phi(\x,t)$ itself in the OJK theory,
and by its correlations in our approach.
We have checked that, despite this difference, anomalous aging
(Eq.~(\ref{th})) applies exactly to the OJK theory with drift,
again with the simultaneous validity of Eq.~(\ref{iden}).

Our last example relates to the smoothing of sandpile surfaces~\cite{coupled}
represented in earlier work by noisy coupled
equations for the fields $h(\x,t)$ and $\rho(\x,t)$.
Here $h(\x,t)$ represents the local height fluctuations of the sandpile surface
( caused by clusters of `stuck' grains) defined with respect to a mean surface,
while $\rho(\x,t)$
represents the local density of mobile grains rolling down the
clusters~\cite{mln}.
The simplest example of these coupled equations is fully linear:
\be
\left\{
\begin{array}{l}
\frah{\p h}{\p t}=D_h\nabla^2h+\c\cdot\nabla h+\eta(\x,t),\\
\frah{\p\rho}{\p t}=D_\rho\nabla^2\rho-\c\cdot\nabla h,
\end{array}
\right.
\label{hr}
\ee
where $\eta(\x,t)$ is Gaussian white noise, such that
$\mean{\eta(\x,t)\eta(\x',t')}=\Delta^2\delta^{(d)}(\x-\x')\delta(t-t')$.
The equation for $h(\x,t)$ is nothing but the Edwards-Wilkinson (EW)
equation~\cite{EW} with a flow (drift) term added;
in the context of sandpiles, this represents the effect of tilt~\cite{mln}.

In Ref.~\cite{coupled} it was found that
while $S_h(\k,t)$, a one-time quantity,
shows only the smoothing exponents of the EW equation
(i.e., the flow term can indeed be Galilean transformed away in this case),
$S_h(\k,\omega)$, a quantity which involves an integration over many times,
shows a clear crossover from EW exponents to a {\it smoothing}
state with $\alpha=\beta=0$, when the drift velocity is large enough.
Evaluating the two-time correlation function of the $h$-field, we find
\be
C_h(t,s)
=\mean{h(\0,t)h(\0,s)}
=\Delta^2\int_0^s\d u\,\big(4\pi D_h(t+s-2u)\big)^{-d/2}
\exp\left(-\frac{c^2(t-s)^2}{4D_h(t+s-2u)}\right).
\ee
In the scaling regime, where both $s$ and $t-s=\tau$ are large
compared to the intrinsic time scale $\tau_0=4D_h/c^2$,
the integral is dominated by microscopic times $u\sim\tau_0$.
We thus recover the anomalous aging result~(\ref{th}),
with $D_1=2D_h$, $D_2=D_h$, so that Eq.~(\ref{iden}) again holds.

A corollary of this result is that anomalous aging
(or, equivalently, the anomalous smoothing referred to
in Ref.~\cite{coupled}) will manifest itself at experimentally
observable times only for large~$c$ -- this, in the sandpile context,
would correspond to unstable packings
of grains susceptible to rolling under tilt~\cite{mln}.
It would be interesting to see if measurements on such a surface indicate
that it is composed of large domains characterized by
EW roughening exponents stacked end on end.
Physically, this is one possible realization
of the anomalous smoothing scenario -- {\em inter-domain} roughness
measurements would here show no significant variation in roughness,
while {\em intra-domain} measurements would show EW roughening exponents.

To sum up, we have demonstrated that a uniform drift velocity
is relevant for the long-time dynamics of systems far from equilibrium.
We have shown that the prediction~(\ref{th})
of a simple linear theory, referred to as anomalous aging,
provides an accurate or even exact description of the non-equilibrium
dynamics of a wide variety of systems, including the oscillatory phase
observed recently in a model of competitive learning, the voter model,
the Ohta-Jasnow-Kawasaki theory for domain growth,
and a theory for the smoothing of sandpile surfaces.

\acknowledgments

It is a pleasure for us to thank L. Berthier, L. Cugliandolo, S. Franz,
and M. Laessig for fruitful discussions.

\end{document}